\documentclass{extarticle}
\usepackage{fullpage}
\usepackage[dvipsnames]{xcolor}
\usepackage{booktabs}
\usepackage{enumitem}
\usepackage { array }
\usepackage { tabularx }
\usepackage { calc }
\usepackage{amsthm}
\usepackage{amsmath}
\usepackage{amsfonts}
\usepackage{graphicx}
\usepackage{hyperref}
\usepackage{soul}

\makeatletter
  {\par}
\makeatother

\let\OLDthebibliography\thebibliography
\renewcommand\thebibliography[1]{
  \OLDthebibliography{#1}
  \setlength{\parskip}{0pt}
  \setlength{\itemsep}{0pt plus 0.3ex}
}

\let\olditem\item
\renewcommand{\item}{%
\olditem\vspace{-2pt}}

\begin{document}
\title{Prediction-Based Decisions and Fairness:\\ A Catalogue of Choices, Assumptions, and Definitions}
\author{Shira Mitchell\\
Civis Analytics\\
sam942@mail.harvard.edu\\
\and
Eric Potash\\
University of Chicago\\
epotash@uchicago.edu\\
\and
Solon Barocas\\
Microsoft Research and Cornell University\\
sbarocas@cornell.edu \\
\and
Alexander D'Amour \\
Google Research \\
alexdamour@google.com \\
\and
Kristian Lum \\
University of Pennsylvania \\
kl1@seas.upenn.edu
}

\maketitle

\begin{abstract}
A recent flurry of research activity has attempted to quantitatively define ``fairness" for decisions based on statistical and machine learning (ML) predictions. The rapid growth of this new field has led to wildly inconsistent terminology and notation, presenting a serious challenge for cataloguing and comparing definitions. This paper attempts to bring much-needed order.

First, we explicate the various choices and assumptions made---often implicitly---to justify the use of prediction-based decisions. Next, we show how such choices and assumptions can raise concerns about fairness and we present a notationally consistent catalogue of fairness definitions from the ML literature. In doing so, we offer a concise reference for thinking through the choices, assumptions, and fairness considerations of prediction-based decision systems.
\end{abstract}

\section{Introduction}

Prediction-based decision-making has swept through industry and is quickly making its way into government. These techniques are already common in lending \cite{Hardt2016,Liu2018,Fuster2018}, hiring \cite{Miller2015a, Miller2015b, HuChen2018b}, and online advertising \cite{Sweeney2013}, and increasingly figure into decisions regarding pretrial detention \cite{Angwin2016,Larson2016,Dieterich2016}, immigration detention \cite{Koulish2016}, child maltreatment screening \cite{Vaithianathan2013, Eubanks2018, Chouldechova2018}, public health \cite{MODA_cooling_towers, potash2015predictive}, and welfare eligibility \cite{Eubanks2018}. Across these domains, decisions are based on predictions of an outcome deemed relevant to the decision. In recent years, attention has focused on how consequential prediction models may be ``biased"-- a now overloaded word that in popular media has come to mean that the model's predictive performance (however defined) unjustifiably differs across disadvantaged groups along social axes such as race, gender, and class.  
Uncovering and rectifying such ``biases" via alterations to standard statistical and machine learning models has motivated a field of research we will call Fairness in Machine Learning (ML).

Fairness in ML has been explored in popular books \cite{ONeil2016, Eubanks2018} and a White House report \cite{Executive2016}, surveyed in technical review papers \cite{Berk2017, Friedler2018, ChouldechovaRoth2018, VermaRubin2018}, 
and an in-progress textbook \cite{barocas-hardt-narayanan}, and inspired a number of software packages \cite{Zehlike2017, Aequitas, angell2018themis,galhotra2017fairness,GoogleWhatIf}. Though the ML fairness conversation is somewhat new, it resembles older work. For example, since the 1960s, psychometricians have studied the fairness of educational tests based on their ability to predict performance (at school or work) \cite{Cleary1966,Thorndike1971,Darlington1971,EinhornBass1971,PetersenNovick1976,HunterSchmidt1976,Lewis1978,HutchinsonMitchell2019}. More recently, Dorans and Cook review broader notions of fairness, including in test design and administration \cite{Dorans2016}.

Importantly, the Fairness in ML field is not purely mathematical. Any definition of fairness necessarily encodes social goals in mathematical formalism. Thus, the formalism itself is not meaningful outside of the particular social context in which a model is built and deployed. For this reason, the goal of this paper is not to advance axiomatic definitions of fairness, but to summarize the definitions and results in this area that have been formalized to date. Our hope with this paper is to contribute a concise, cautious, and reasonably comprehensive catalogue of the important fairness-relevant choices that are made in designing a predictive model, the assumptions that underlie many models where fairness is a concern, and some metrics and methods for evaluating the fairness of models. Alongside this summary, we point out gaps between mathematically convenient formalism and the larger social goals that many of these concepts were introduced to address.

Throughout this article, we ground our theoretical and conceptual discussion of Fairness in ML in real-world example cases that are prevalent in the literature: pretrial risk assessment and lending models. In a typical pretrial risk assessment model, information about a person who has been arrested is used to predict whether they will commit a (violent) crime in the future or whether they will fail to appear for court if released. These predictions are often based on demographic information, the individual's criminal history, and sometimes their responses to more in-depth interview questions. These predictions are then used to inform a judge, who must make an extremely consequential decision: whether the person who has been arrested should be released before their case has concluded, and if so, under what conditions. Although the options available to the judge or magistrate are many, including simply releasing the person, setting bail, requiring participation in a supervised release program, etc., in this literature, this decision is often reduced to a binary decision: release or detain. In lending, the task is to predict whether an individual will repay a loan if one is granted. These predictions are based on employment, credit history, and other covariates. The loan officer then incorporates this prediction into their decision-making to decide whether the applicant should be granted a loan, and if so, under what terms. In the Fairness in ML literature, the decision space is also often reduced to the decision to grant or deny the loan.

The paper proceeds as follows. Section \ref{preliminaries} outlines common choices and assumptions that are made that are often considered outside the scope of the model but have material consequences for the fairness of the model's performance in practice. In Section \ref{setup}, as a segue to the slightly more mathematical parts of the paper, we introduce our main setup and notation. Section \ref{flavors} begins the discussion of the various mathematical notions of fairness that have been introduced, including tensions and impossibilities among them. In Section \ref{causal_fairness} we explore causal frameworks for reasoning about algorithmic fairness. Section \ref{ways_forward} offers some suggestions and ideas for future work in this area, and Section \ref{conclusion} concludes.

\section{Choices, assumptions, and considerations}\label{preliminaries}
Several recent papers have demonstrated how social goals are, sometimes clumsily, abstracted and formulated to fit into a prediction task \cite{danks2017algorithmic, silva2018algorithms, Green2018, GreenHu2018, Ochigame2018, Dobbe2018, Selbst2018,  eckhouse2018layers, passi2019problem}. In this section, we link these socio-technical concerns to choices and assumptions made in the policy design process. Broadly, these assumptions take the problem of evaluating the desirability of a policy, and reduce it to the simpler problem of evaluating the characteristics of a model that predicts a single outcome.

\subsection{The policy question}
Much of the technical discussion in Fairness in ML takes as given the social objective of deploying a model, the set of individuals subject to the decision, and the decision space available to decision-makers who will interact with the model's predictions. Each of these are {\it choices} that---although sometimes prescribed by policies or people external to the immediate model building process---are fundamental to whether the model will ultimately advance fairness in society, however defined. 

\subsubsection{The over-arching goal}\label{goal}
Models for which fairness is a concern are typically deployed in the service of achieving some larger goal. For a benevolent social planner, this may be some notion of justice or social welfare \cite{HuChen2018a}. For a criminal justice actor, this goal may be reducing the number of people who are detained before their trial while simultaneously protecting public safety. For a bank making lending decisions, the goal may be maximizing profits. Often there are genuinely different and conflicting goals, which are not resolved by more data \cite{Eubanks2018,ONeilGunn2018}. If one disagrees with the over-arching goal, a model that successfully advances this goal---regardless of whether it attains any of the mathematical notions of fairness discussed here---will not be acceptable \cite{passi2019problem}.

Prediction-based decision systems often implicitly assume the pursuit of the over-arching social goal will be served by better predicting some small number of outcomes. For example, in pretrial decisions, outcomes of interest are typically  crime (measured as arrest or arrest for a violent crime) and non-appearance in court. In contrast, human decision-makers may consider several outcomes, including impacts on a defendant's well-being or caretaker status \cite{S770}. In making decisions about college admissions, it may be tempting to narrow the larger goals of higher education to simply the future GPAs of admitted students \cite{Kleinberg2018}. Narrowing focus to a chosen, measured outcome can fall short of larger goals (this is sometimes called {\it omitted payoff bias} \cite{kleinberg2017human}).

Furthermore, prediction-based decision systems usually only focus on outcomes under one decision (e.g. crime if released) and assume the outcome under an alternative decision is known (e.g. crime if detained). More recent work has formalized this oversight using the language of counterfactuals and potential outcomes \cite{coston2020counterfactual}. Finally, prediction-based decisions are formulated by assuming progress towards the over-arching goal can be expressed as a scalar utility function that depends only on decisions and outcomes, see Section \ref{setup}.

\subsubsection{The population} 
A model's predictions are not applied to all people, but typically to a specific sub-population. In some cases, individuals choose to enter this population; in other cases, they do not. In pretrial decisions, the population is people who have been arrested. In lending decisions, the population is loan applicants. These populations are sampled from a larger population by some mechanism, e.g. people are arrested because a police officer determines that the individual's observed or reported behavior is sufficiently unlawful to warrant an arrest; creditors target potential applicants with offers or applicants independently decide to apply to a particular lending company for a loan.  The mechanism of entry into this population may reflect objectionable social structures, e.g. policing that targets racial minorities for arrest \cite{Alexander2012} and discrimination in loan pre-application screening \cite{Courchane2000}. A model that satisfies fairness criteria when evaluated only on the population to which the model is applied may overlook unfairness in the process by which individuals came to be subject to the model in the first place. 

\subsubsection{The decision space}\label{decision_space} The decision space is the set of actions available to a decision maker. For example, in a simplified pre-trial context, the decision space might consist of three options: release the arrested person on recognizance, set bail that must be paid to secure the individual's release, or detain the individual. In the lending example, the decision space might only consist of the options to grant or deny the loan application. Both of these decision spaces leave out many other possible interventions or options. For example, a different pre-trial decision space might include the option that the judge recommend the person who has been arrested make use of expanded pretrial services \cite{koepke2018danger, vannostrand2011state, mahoney2004pretrial}, including offering funds to help defendants with transportation to court, paying for childcare, text message reminders of court dates \cite{courtmessaging, cooke2018using, uptrusttext, NYCtext}, drug counseling, or locating job opportunities. In lending, a broader decision space could include providing longer-term, lower-interest loans. In several contexts, one could consider replacing individual-level with community-level policies. While mathematical definitions of the ML fairness literature discussed below may be able to certify the fair allocation of decisions across a population, they have nothing to say about whether any of the available actions are acceptable in the first place.

\subsection{The statistical learning problem}\label{learning problem}
Mathematical definitions of fairness generally treat the statistical learning problem that is used to formulate a predictive model as external to the fairness evaluation.
Here, too, there are a number of choices that can have larger social implications, but go unmeasured by the fairness evaluation.
We focus specifically on the choices of training data, model, and predictive evaluation.

\subsubsection{The data}
\begin{figure}[h!]
\center
\includegraphics[scale=.25]{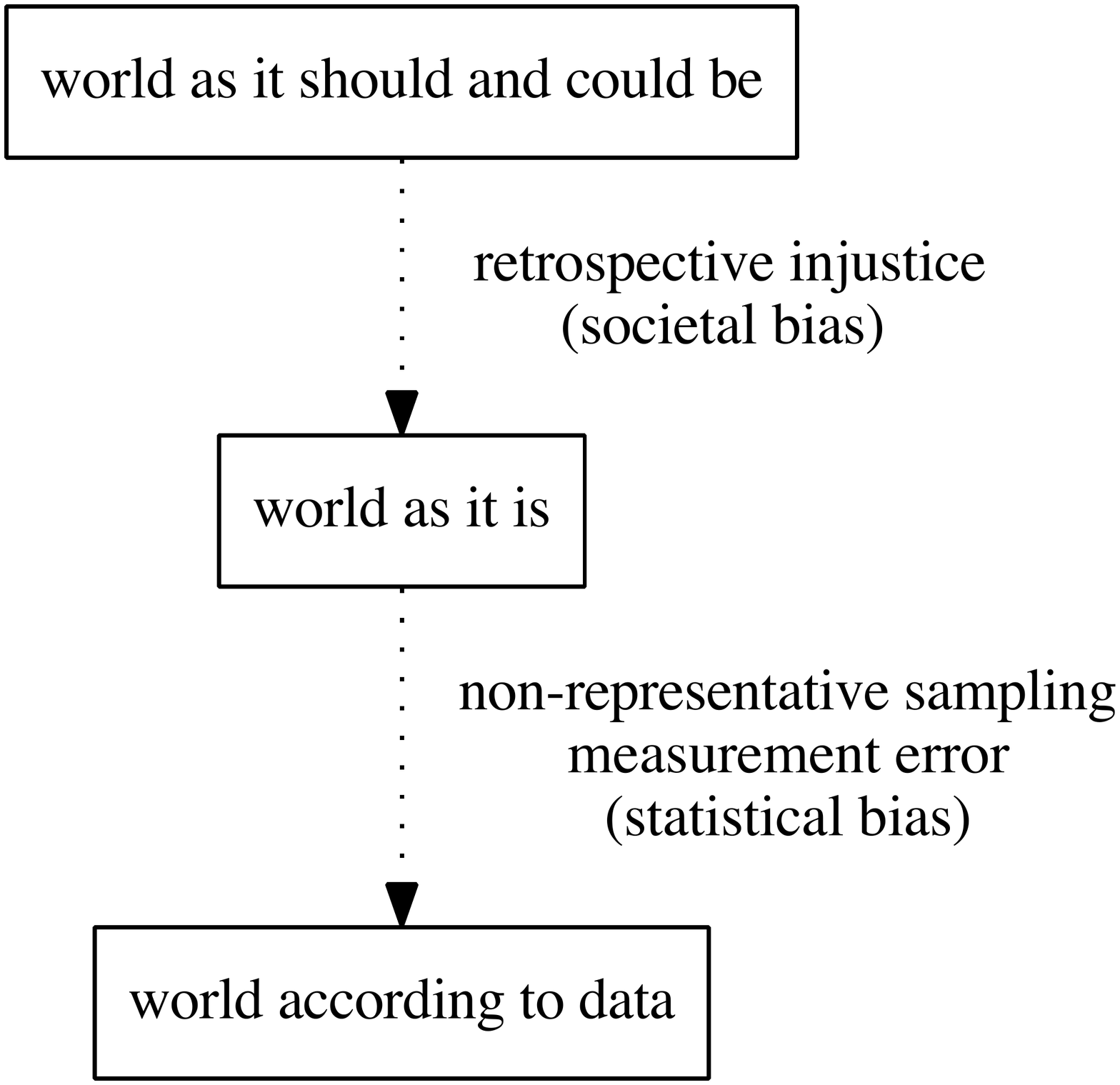}
\caption{A cartoon showing two components of ``biased data": societal bias and statistical bias.}\label{biased_data}
\end{figure}

The foundation of a predictive model is the data on which it is trained.
The data available for any consequential prediction task, especially data measuring and categorizing people, can induce undesirable properties when used as a basis for decision-making.
In the Fairness in ML literature, data with these undesirable properties are often labeled informally as ``biased" (e.g., \cite{KamiranCalders2009, BarocasSelbst2016, Chouldechova2017, Lipton2018}).
Here, we decompose this notion of bias into more precise notions of \emph{statistical} bias---i.e. concerns about non-representative sampling and measurement error---and \emph{societal bias}---i.e. concerns about objectionable social structures that are represented in the data (figure~\ref{biased_data}).
We treat each of these notions in turn.

\paragraph{Statistical bias}
Here we consider statistical bias to be a systematic mismatch between the sample used to train a predictive model, and the world as it currently is.
Specifically, we consider how sampling bias and measurement errror can induce fairness implications that are usually unmeasured by mathematical fairness definitions.

\emph{Sampling bias} occurs when a dataset is not representative of the full population to which the resulting model will be applied. For example, in pretrial decisions, the sample used to train the model is typically drawn from the population of people who were released pre-trial. For people who were not released before their case has concluded, it is not possible to directly measure typical prediction outcomes, like re-arrest, as they do not have the opportunity to be re-arrested. These people are typically excluded from training data sets. In the lending example, models may be based only on those individuals who were granted a loan, as it is not possible to measure the outcome variable--  whether they would have repaid the loan had it been granted.
This problem is sometimes called {\it selective labels} \cite{kleinberg2017human, lakkaraju2017selective,Chouldechova2018,CorbettGoel2018,De-Arteaga2018}. Incorrectly assuming the sample is representative can lead to biased estimation of conditional probabilities (e.g. probability of crime given covariates), biased estimation of utility, and inadequate fairness adjustments \cite{kallus2018residual}.

Under a missing at random assumption, 
modeling could hope to avoid this selection bias \cite{Gelman2013}. But there can be regions of the covariates where no data exists to fit a model; for there may be values of a covariate for which no defendants were ever released 
and hence the outcome in that region is unobserved \cite{ChouldechovaRoth2018}. One option is extrapolation: fit the model to released defendants, 
then apply the model to all defendants even if this includes new regions of the variables. However, ML models can perform poorly with such a shift in covariates \cite{Rabanser2018}. Another option is to trust previous decisions: assume that regions of the covariates where no defendants were released are regions where all defendants would have committed crimes if released \cite{De-Arteaga2018}. Though convenient from a modeling perspective, these types of assumptions are unlikely to hold.

\label{measurement}
Similarly \emph{systematic measurement error}, particularly when the error is greater for some groups than others (known as {\it differential measurement error} \cite{Vanderweele2012}), can have profound consequences. In lending, some measures of past success in loan repayment (measures that may be used to predict future loan repayment) only account for repayment of loans through formal banking institutions. At least historically, immigrant communities were more likely to engage in more informal lending, and so measurements of past success in loan repayment may systematically understate the value of past loans repaid for people who participate in informal lending relative to those who go through formal channels \cite{day2002credit}. Similar issues exist in our running example of pretrial risk assessment, where commission of a crime is often measured by re-arrest during the pre-trial period. Individuals and groups that are more likely to be re-arrested following the commission of a crime will be measured in the data as more criminal, regardless of whether those differences are the result in true underlying rates of the commission of the crime or bias in the process by which the decision to arrest is made. 

Additionally, the perceived fairness of a model may hinge on measurement choices that incorporate moral or normative arguments.  For example, in one author's experience, recent updates to a pre-trial risk assessment tool have incorporated changes such  that past failures to appear for court appointments  have a sunset window for inclusion in a model. In the name of fairness, after the sunset window, they can no longer be ``counted against" the defendant. In lending, it is already the case that bankruptcy is erased from one's credit report after seven (chapter 13 bankruptcy) or 10 years (chapter 7 bankruptcy). These measurement considerations have less to do with the accuracy or bias of the measurement and more to do with normative decisions about how a person ought to be evaluated.

\paragraph{Societal bias}
Even if training data are representative and accurate, they may still record objectionable social structures that, if encoded in a policy, run counter to the decision-maker's goals. This corresponds to a non-statistical notion of \emph{societal bias} \cite{suresh2019framework}.
There is overlap between the two concepts, e.g. using arrests as a measure of crime can introduce statistical bias from measurement error that is differential by race because of a racist policing system \cite{Alexander2012, LumIsaac2016}.
But suppose we could perfectly measure crime, does this make the data free from ``bias"? In a statistical sense, yes.\footnote{In statistics, ``bias" refers to properties of an estimator, not data. Here we mean bias in the estimation of conditional probabilities or fairness metrics that could result from non-representative data, measurement error, or model misspecification.} In a normative sense, no, because crime rates reflect societal bias, including how crime is defined \cite{nationcriminalize}.
In general, addressing issues of societal bias may require adjusting data collection processes, or may not have technical solutions at all.

\subsubsection{The predictive model}
Statistical and machine learning models are designed to identify patterns in the data used to train them. As such, they will reproduce, to the extent that they are predictive, unfair patterns encoded in the data or unfairness in the problem formulation described above.

Furthermore, there are many choices for building a model. Indeed, this is the subject of much of statistics and machine learning research. Choices include a class of model, a functional form, and model parameters, among others. These choices, too, have ramifications for fairness. For example, it has been shown that different functional forms can differ in their estimates for individuals and disparities between groups of individuals \cite{chouldechova2017fairer}. 

Models also differ in their interpretability. For example, some use point systems, where the presence of absence of characteristics are assigned integer point values. The individual's final score or prediction is then the sum of those integer-valued points. These and other simple models can facilitate ease of understanding of how perturbations to the inputs might impact the model's predictions. Human-interpretable models allows insight into cases in which the model will produce counter-intuitive or non-sensical results, and can help humans catch errors when they occur \cite{Rudin2018PleaseSE}.

Finally, there is the choice of {\it which} covariates to include in the model. Outcome predictions can change depending on what we condition on. Person A can have a higher predicted value than Person B with a choice of covariates $V$, but a lower predicted value with a choice of covariates $V'$. 

\subsubsection{Evaluation}\label{evaluation}
Fairness concerns aside, predictive models are typically built, evaluated, and selected using various measures of their predictive performance such as (for continuous predictions) mean squared error or (for binary predictions) positive-predictive value, sensitivity, specificity, etc.

We note that such evaluations generally make three important assumptions. First, they assume that decisions can be evaluated as an aggregation of {\it separately} evaluated individual decisions. This includes assuming that outcomes are not affected by the decisions for others, an assumption known as {\it no interference} \cite{ImbensRubin2015}. In the loan setting, denying one family member a loan may directly impact another family member's ability to repay their own loan. If both are evaluated by the same model, this dependence is in conflict with the ``separately" assumption.

This assumption resembles beliefs from utilitarianism, which represents social welfare as a sum of individual utilities \cite{Rawls}. With binary predictions and decisions, each individual decision's utility can in turn be expressed in terms of four utilities for each possibility of true positive, false positive, false negative, and true negative.

Second, these evaluations assume that all individuals can be considered \textit{symmetrically}, i.e. identically. This assumes, for example, that the harm of denying a loan to someone who could repay is equal across people. Denying someone a loan for education will likely have a very different impact on their life than denying a different person a similarly sized loan for a vacation home.

Third, these evaluations assume that decisions are evaluated \textit{simultaneously}. That is, they are evaluated in a batch (as opposed to serially \cite{ChouldechovaRoth2018}) and so do not consider potentially important temporal dynamics. For example, Harcourt shows that if the population changes their behavior in response to changing decision probabilities, prediction-based decisions can backfire and diminish social welfare \cite{harcourt2008against}. 

A fundamental question of Fairness in ML is to what extent prediction measures making these assumptions are relevant to fairness.

\subsection{Axes of fairness and protected groups} \label{groups}
A final choice in mathematical formulations of fairness is the axis (or axes) along which fairness is measured.
For example, much of the ML fairness literature considers the simple case of two groups (advantaged and disadvantaged).
In this setting, typically only one variable is selected as a ``sensitive attribute," which defines the mapping to the advantaged and disadvantaged groups. Deciding how attributes map individuals to these groups is important and highly context-specific. Does race determine advantaged group membership? Does gender? In regulated domains such as employment, credit, and housing, these so-called ``protected characteristics" are specified in the relevant discrimination laws. Even in the absence of formal regulation, though, certain attributes might be viewed as sensitive, given specific histories of oppression, the task at hand, and the context of a model's use. Differential treatment or outcomes by race is often concerning, even when it does not violate a specific law, but which racial groups are salient will often vary by cultural context.

Though most Fairness in ML work considers only one sensitive attribute at a time, discrimination might affect members at the intersection of two groups, even if neither group experiences discrimination in isolation \cite{Crenshaw1989}. This fact was most famously highlighted in the influential work of Crenshaw, who analyzed failed employment discrimination lawsuits involving black women who could only seek recourse against discrimination as black women which they were unable to establish simply as sex discrimination (since it does not apply to white women) or as race discrimination (since it does not apply to black men) \cite{Crenshaw1989}. More recently, the importance of considering combinations of attributes to define advantage and disadvantage in Fairness in ML applications is seen in \cite{BuolamwiniGebru2018}, which  evaluated commercial gender classification systems and found that darker-skinned females are the most misclassified group.

\section{Setup and notation}\label{setup}
We now turn to mathematical formulations of fairness, having presented a number of key choices, assumptions, and consideration that make this abstraction possible.
Here, we introduce notation for some canonical problem formulations considered in the Fairness in ML literature.
We follow much of the recent discourse in this literature and focus on fairness in the context of binary decisions that are made on the basis of predictions of binary outcomes.

We consider a {\it population} about whom we want to make decisions. We index people by $i=1,...,n$. We assume this finite population (of size $n$) is large enough to approximate the ``superpopulation'' \cite{Gelman2013} distribution from which they were drawn, and refer to both as the ``population". Each person has covariates (i.e. features, variables, or inputs) $v_i \in \mathcal{V}$ that are known at decision time. In some cases, we can separate these into sensitive variable(s) $a_i$ (e.g. race, gender, or class) and other variables $x_i$, writing $v_i = (a_i,x_i)$.

A binary decision $d_i$ is made for each person. We restrict decisions to be functions of variables known at decision time, $\delta: \mathcal{V} \rightarrow \{0,1\}$, where $d_i = \delta(v_i)$.
We define random variables $V$, $Y$, $D = \delta(V)$ as the values of a person randomly drawn from the population. 

In {\it prediction-based decisions}, decisions are made based on a prediction of an outcome, $y_i$, that is unknown at decision time. Specifically, decisions are made by first estimating the conditional probability
$$P[Y=1|V=v_i].$$

The decision system does not know the true conditional probability; instead it uses $\psi: \mathcal{V} \rightarrow [0, 1]$ where $s_i = \psi(v_i)$ is a score intended
to estimate $P[Y = 1|V = v_i]$. Let $S = \psi(V)$ be a random score from the population. A prediction-based decision system then has a decision function $\delta$ that is a function of the score alone, i.e. $\delta(v) = f(\psi(v))$ for some function $f$. Both $\psi$ and $\delta$ are functions of a sample of $\{(v_i, y_i)\}$ that we hope resembles the population.

For example, in pre-trial risk assessment we predict whether an individual $i$ will be re-arrested ($y_i = 1$) or not ($y_i = 0$) using $v_i$, summaries of criminal history and basic demographic information. A decision is then made to detain ($d_i = 1$) or release ($d_i = 0$) the individual on the basis of the prediction. In the lending setting, the goal is to predict whether an individual will default on ($y_i = 0$) or repay ($y_i = 1$) as a function of $v_i$, their credit history. The decision space consists only of the decision to deny ($d_i = 0$) or grant ($d_i= 1$) the loan. In both examples, we choose notation such that when $y_i = d_i$, the ``correct" decision has been made. 

\section{Flavors of fairness definitions from data alone}\label{flavors}
We begin our exposition of formal fairness definitions with so-called ``oblivious'' definitions of fairness \cite{Hardt2016} that depend on the observed data alone.
These definitions equate fairness with certain parities that can be derived from the distributions of the observed features $V$, outcomes $Y$, scores $S$, and decisions $D$, without reference to additional structure or context.
These stand in contrast to non-oblivious fairness definitions, presented in Section~\ref{causal_fairness}.

\subsection{Unconstrained utility maximization and single-threshold fairness} \label{default}
As a default, we consider a definition of fairness, which we call {\it single-threshold fairness}, that is fully compatible with simply maximizing a specific kind of utility function without treating fairness as a separate consideration. 
Here, a decision is considered to be fair if individuals with the same score $s_i = \psi(v_i)$ are treated equally, regardless of group membership \cite{CorbettGoel2018}.

This notion of fairness is connected to utility maximization by a set of results showing that, for a certain set of utility functions and scores $\psi(v)$, the utility-maximizing decision rule $\delta$ is necessarily a {\it single-threshold rule} \cite{Karlin1956,Berger1985,Corbett2017, Lipton2018}. 
Rules of this form select a threshold $c$ and apply one decision to individuals with scores below $c$ and another to individuals with scores above $c$. Formally we have
\[\delta(v) = I(\psi(v) \geq c).\]
For example, if the outcome is loan repayment, individuals with scores above the threshold would be granted a loan while those with scores below would be denied.
A key condition for the optimality of single-threshold rules is that the score $\psi(v)$ is a good approximation the true conditional probability $P[Y=1|V=v]$.

Moreover, since the optimal threshold depends only on the utility function, the single-threshold rule maximizes utility within any subgroup as well. In this sense the single-threshold rule is a viable group-sensitive definition of fairness, as it is optimal for both groups under the outlined assumptions.

The desirability of single-threshold rules is sensitive to a number of the choices outlined in Section~\ref{preliminaries}.
First, these rules are a direct function of the score $\psi(v)$ and the utility function used to evaluate the decision $\delta(v)$.
These functions are specified by the decision-maker, and are sensitive to data collection, measurement, and modeling choices (Section~\ref{learning problem}).
In addition, the optimality results here make strong assumptions about the form of the utility function used to evaluate the decision $\delta(v)$.
In particular, they only hold for utility functions which satisfy the separate, symmetric, and simultaneous assumptions of Section~\ref{evaluation}.

These sensitivities motivate notions of fairness that are external to the utility maximization problem, and which can be evaluated without taking scoring models or utility functions for granted. 

\subsection{Equal prediction measures}\label{predictions_fairness}
If the impacts of decisions are contained within groups, their utilities can be considered group-specific. A notion of fairness might ask these to be equal.

When false positives and false negatives have equal cost, this corresponds to the fairness definition of {\bf Equal accuracy}: $P[D=Y| A=a] = P[D=Y | A=a']$. This definition is based on the understanding that fairness is embodied by the predictions being ``correct" at the same rate among groups.  For example, we might want a medical diagnostic tool to be equally accurate for people of all races or genders.

\begin{figure}[h]
\centering
\includegraphics[height=2in]{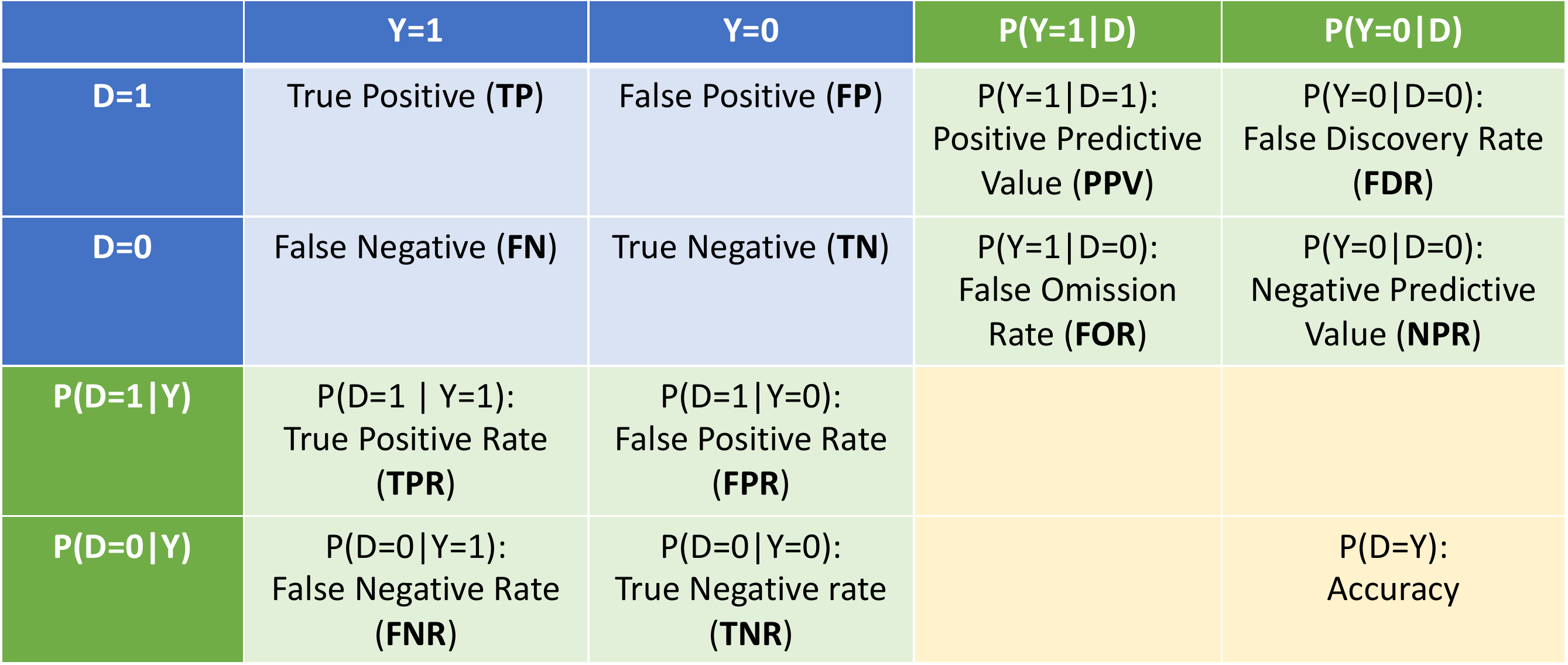}
\caption{\label{confusion} Confusion matrix defining terminology used in this article for relationships between $Y$ and $D$.}
\end{figure}

Instead of comparing overall accuracies, we could restrict the comparison to subsets of our advantaged and disadvantaged groups defined by their predictions or their outcomes.  All such possible subsets are summarized by a {\it confusion matrix}, which illustrates match and mismatch between $Y$ and $D$, with margins expressing conditioning on subsets; see Figure \ref{confusion}, which defines common terminology for quantities contained in this table that will be discussed in this manuscript.

\subsubsection{Definitions from the confusion matrix}

For any box in the confusion matrix involving the decision $D$, we can require equality across groups. For example, we could define fairness by Equality of False Positive Rates by requiring that the model satisfy $P[D=1 | Y=0, A=a] = P[D=1 | Y=0, A=a']$. All other cells of the confusion matrix can similarly define fairness analogously by adding the conditioning on sensitive group attribute, $A$. We list common definitions of fairness that have been proposed in the literature from the margins of the confusion matrix, grouped by pairs that sum to one. Equality of one member of the pair immediately implies equality of the other.

\paragraph{Conditional on outcome}
First consider conditioning on the outcome $Y$. This leads to two pairs of fairness definitions. The first pair, {\bf Equality of False Positive Rates} and {\bf Equality of True Negative Rates}, conditions on $Y=0$ and is equivalent to requiring $D \perp A \ |\  Y=0$. This definition of fairness  demands equality from the perspective of those individuals with the outcome defined by zero. For example, this reflects {\it the perspective of ``innocent" defendants}, in requiring that all individuals who do not go on to be re-arrested have the same likelihood of being released, regardless of whether they are in the advantaged or disadvantaged group \cite{Narayanan2018}.

The second pair, {\bf Equality of True Positive Rates} and {\bf Equality of False Negative Rates}, is equivalent to $D \perp A \ |\  Y=1$. This definition of fairness takes the perspective of those individuals with outcome defined by one. For example, this is equivalent to the requirement that all individuals who will go on to repay a loan have the same likelihood of receiving a loan, regardless of their sensitive group membership. This condition alone has been called {\it equal opportunity} \cite{Hardt2016}. Taken together with Equality of FPR/TNR, these notions of fairness are called {\it error rate balance} \cite{Chouldechova2017}, {\it separation} \cite{barocas-hardt-narayanan} or {\it equalized odds} \cite{Hardt2016}).

These two pairs reflect a fairness notion that people with the same outcome are treated the same, regardless of sensitive group membership. We posit that this notion of fairness is more closely aligned with the perspective of the population evaluated by the model as it demands that people who are {\it actually} similar (with respect to their outcomes) be treated similarly. \cite{Hardt2016} gives an algorithm for model fitting that is optimal with respect to this notion of fairness. 

\paragraph{Conditional on decision}
Turning to the other margin of the confusion matrix, {\bf Equality of Negative Predictive Value} and {\bf Equality of Negative Predictive Value} are defined by the statement $Y \perp A \ |\  D=0$. This notion of fairness conditions on having received the decision defined by zero. For example, this definition requires that all individuals who were granted a loan were equally likely to have defaulted. The other pair of definitions that appear on this margin are {\bf Equality of Positive Predictive Value} and {\bf Equality of False Discovery Rate}. This is defined by the conditional independence relationship $Y \perp A \ |\  D=1$. This definition of fairness is also sometimes called called {\it predictive parity} \cite{Chouldechova2017}, and it is assessed by an {\it outcome test} \cite{Simoiu2017}. This definition of fairness, for example, is met when people from both the advantaged and disadvantaged groups who are denied loans would have repay them at the same rate. These two pairs of definitions taken together have been called {\bf sufficiency} \cite{barocas-hardt-narayanan}. They reflect a fairness notion that people with the same decision would have had similar outcomes, regardless of group.

These two pairs of definitions of fairness reflect the viewpoint of the decision-maker or modeler, as individuals are grouped with respect to the decision or model's prediction, not with respect to their actual outcome. \cite{dieterich2016compas} argues that this definition of fairness is more appropriate because at the time of the decision, the decision-maker knows only the prediction, not the eventual outcome for the individual, and so individuals should be grouped by the characteristic that is known at the time of the decision. 

Zafar et al. call all four pairs of definitions (as well as equal accuracy) {\it avoiding disparate mistreatment} \cite{Zafar2017b}. Berk et al. \cite{Berk2017} also consider a definition based on several of the above elements of the confusion matrix. They define {\bf Treatment Equality} to be  the ratio of false negatives to false positives: $\frac{P[Y=1, D=0 | A=a]}{P[Y=0, D=1 | A=a]} = \frac{P[Y=1, D=0 | A=a']}{P[Y=0, D=1 | A=a']}$.

\subsubsection{Analogues with scores}\label{Scores}
In the previous section, we focused on fairness definitions defined by the relationship between $Y$ and $D$, a binary decision. As discussed in Section \ref{default}, the ultimate decision is typically arrived at by thresholding a score $S = \psi(V)$ that is intended to estimate $P[Y=1|V=v]$. In this section, we consider definitions based on the relationship between $S$ and $Y$ and draw connections to the confusion matrix-based definitions. 

{\bf AUC parity} is the requirement that the area under the receiver operating characteristic (ROC) curve is the same across groups. This is analogous to Equality of Accuracy in the previous section, as the AUC is a measure of model accuracy.

{\bf Balance for the Negative Class} is defined by $E(S | Y = 0, A = a) = E(S | Y = 0, A = a')$. This is similar to Equality of False Positive rates in that if the score function is the same as the binary decision, i.e. $S=D$, then achieving Equality of False Positive rates implies Balance for the Negative Class. This is easily seen because the conditional expectation of a binary variable is, by definition, the same as the conditional probability that variable takes value one. {\bf Balance for the Positive Class} is similarly defined as $E(S | Y = 1, A = a) = E(S | Y = 1, A = a')$. By an analogous argument to that above, this definition of fairness is closely related to Equality of True Positive Rates. {\bf Separation} denotes  conditional independence of the protected group variable with the score or decision given $Y$ and covers all definitions discussed in this paragraph.

{\bf Calibration within groups} is satisfied when $P[Y =1| S, A = a] = S$. That is, when the score function accurately reflects the conditional probability of $Y\mid V$. Barocas et al. point out that calibration within groups is satisfied without a fairness-specific effort \cite{barocas-hardt-narayanan}. With enough (representative, well-measured) data and model flexibility, a score $S$ can be very close to $E(Y|A,X)$. With many $X$ variables, $A$ may be well-predicted by them, i.e. there is a function $a(X)$ that is approximately $A$. Then we can get calibration within groups even without using $A$ because $E(Y|A,X) = E(Y|X)$. 

Calibration within groups is the multi-valued analogue of Equality of Positive Predictive Value and Equality of Negative Predictive Value. Sufficiency, $Y \perp A\ |\ D$ or $Y \perp A\ |\ S$, is closely related to both of these cases.  In terms of $S$, calibration within groups implies sufficiency. Conversely, if $S$ satisfies sufficiency then there exists a function $l$ such that $l(S)$ satisfies calibration within groups \cite{barocas-hardt-narayanan}).

\subsection{Equal decision measures}\label{EqualDecisions}

We now turn to fairness notions that focus on decisions $D$ without consideration of $Y$. These can be motivated in a few ways. Suppose that {\it from the perspective of the population about whom we make decisions}, one decision is always preferable to another, regardless of $Y$ (e.g. non-detention, admission into college\footnote{In contrast, lending to someone unable to repay could hurt their credit score \cite{Liu2018}. Of course, the ability to repay may strongly depend on the terms of the loan.}) \cite{Narayanan2018}. In other words, allocation of benefits and harms across groups can be examined by looking at the decision ($D$) alone. Furthermore, while the decisions (e.g. detentions) are observed, the outcome being predicted (e.g. crime if released) may be unobserved or poorly measured, making error rates unknown. Therefore, disparity in decisions (e.g. racial disparity in detention rates) may be more publicly {\it visible} than disparity in error rates (e.g. racial disparity in detention rates among those who would not have committed a crime).

Yet another motivation to consider fairness constraints without the outcome $Y$ is {\it measurement error} (see Section \ref{measurement}). For example, if $Y$ suffers from differential measurement error, fairness constraints based on $Y$ may be unappealing \cite{JohndrowLum2017}. One might believe that all group differences in $Y$ are a result of measurement error, and that the true outcomes on which we want to base decisions are actually similar across groups \cite{Friedler2016}.

Even more broadly, we might consider the relationship between $A$ and $Y$ to be unfair, even if the observed relationship in the data is accurately capturing a real world phenomenon. 
These considerations can all motivate requiring {\bf Demographic Parity}, {\bf Statistical Parity}, {\bf Group Fairness} \cite{Dwork2012}): equal decision rates across groups regardless of outcome $Y$. This fairness definition can be thought of in terms of (unconditional) independence: $S \perp A$ or $D \perp A$. \cite{ kamiran2009classifying, calders2010three, feldman2015certifying, JohndrowLum2017} give algorithms to build models achieving this notion of fairness

A related definition considers parity within strata: {\bf Conditional demographic parity} is defined by the condition that $D \perp A\ |\ \text{Data}$. When $\text{Data} = Y$, conditional demographic parity is equivalent to separation. When $\text{Data} = X$ (the insensitive variables), it is equivalent to {\bf Fairness through Unawareness} \cite{Kusner2017}, {\bf anti-classification} \cite{CorbettGoel2018}, or {\bf treatment parity} \cite{Lipton2018}.  This is easily achieved by not allowing a model to directly access information about $A$. Unawareness implies that people with the same $x$ are treated the same, i.e. $\delta(v_i) = \delta(v_{i'})$ if $x_i=x_{i'}$. Note the opposite is not true, since it is possible that no two people have the same $x$. A related idea requires people who are similar in $x$ to be treated similarly. More generally, we could define a similarity metric between people that is {\it aware} of the sensitive variables, motivating the next flavor of fairness definitions \cite{Dwork2012}.

\subsection{Impossibilities}\label{impossibilities}
Although each flavor of fairness definition presented in this section formalizes an intuitive notion of fairness, these definitions are not mathematically or morally compatible in general.
In this section, we review several impossibility results about definitions of fairness, providing context for how this discussion unfolded in the Fairness in ML literature.

\subsubsection{The COMPAS debate}
In the Fairness in ML literature, incompatibilities between fairness definitions were brought to the fore in a public debate over a tool called COMPAS (Correctional Offender Management Profiling for Alternative Sanctions) deployed in criminal justice settings.

In 2016, ProPublica published a highly influential analysis based on data obtained through public records requests \cite{Angwin2016, Larson2016}. Their most discussed finding was that COMPAS does not satisfy equal FPRs by race: among defendants who did not get rearrested, black defendants were twice as likely to be misclassified as high risk. Based largely on this and other similar findings, they described the tool as ``biased against blacks."

Northpointe (now Equivant), the developers of COMPAS, critiqued ProPublica's work and pointed out that COMPAS satisfies equal PPVs: among those called higher risk, the proportion of defendants who got rearrested is approximately the same regardless of race \cite{Dieterich2016}. COMPAS also satisfies calibration within groups \cite{Flores2016}. Much of the subsequent conversation consisted of either trying to harmonize these definitions of fairness or asserting that one or the other is correct. As it turns out, there can be no harmony among definitions in a world where inequality  and imperfect prediction is the reality.

\subsubsection{Separation and sufficiency}
Tension between margins of the confusion matrix is expressed in three very similar results.
Barocas et al. \cite{barocas-hardt-narayanan} and  
Wasserman \cite{Wasserman2010}\label{sep_suff} show that under the assumption of separation ($S \perp A | Y$) and sufficiency ($Y \perp A | S$), then either $(Y,S) \perp A$ or an event in the joint distribution has probability zero.

Putting this in context and recalling that sufficiency implies equality of PPVs and separation implies equality of FPRs, this result shows that both supported definitions of fairness in the COMPAS debate can only be met when (1) the rate of recidivism and the distribution of scores are the same for all racial groups or (2) there are some groups that never experience some of the outcomes (e.g. white people are never re-arrested). 

A related result was given in Kleinberg et al. \cite{Kleinberg2016}. Here they showed that if a model  satisfies balance for the negative class, balance for the positive class, and calibration within groups, then there are either equal base rates ($Y \perp A$) or there was perfect prediction ($P[Y=1|V=v] = 0$ or 1 for all $v \in \mathcal{V}$). A very similar result was shown by Chouldechova \cite{Chouldechova2017}. In the context of the COMPAS debate, this requires either that reality be fair (i.e. there is no racial disparity in recidivism rates) or the model is perfectly able to predict recidivism (a reality that is as of now unattainable). Equal base rates and perfect prediction can be called trivial, degenerate, or even utopian (representing two very different utopias). Regardless of description, these conditions were not met in ProPublica's data on COMPAS and so the definitions of fairness championed by the different sides of the debate cannot simultaneously be achieved.

\subsubsection{Incompatibilities with demographic parity}
Here we describe several impossibility results involving demographic parity, though they have not played so prominent a role in the public debate about fairness. 
 Barocas et al. \cite{barocas-hardt-narayanan}]\label{sep_demo} showed that when $Y$ is binary and the scoring function exhibits separation ($S \perp A | Y$) and demographic parity ($S \perp A$), then there must be at least one of equal base rates ($Y \perp A$) or a useless predictor ($Y \perp S$). That is, the only way both separation and demographic parity are jointly possible is if WAE is exactly true or the scoring function has no predictive utility for predicting $Y$. 

Similarly, Barocas et al.\cite{barocas-hardt-narayanan}] also showed that if a model satisfies sufficiency ($Y \perp A | S$) and demographic parity ($S \perp A$), then there must also be equal base rates: $Y \perp A$. Taken together, in the context of the COMPAS debate, even if we could decide that Equality of PPVs or Equality of NPVs was the relevant notion of fairness, if we also want to constrain the model to avoid disparate impact by requiring demographic parity, this would only be possible if we lived in a world in which there are no racial disparities in re-arrest and/or we have a completely useless predictive model.

Finally, Corbett-Davies et al. and Lipton et al. both note that a decision rule $\delta$ that maximizes utility under a demographic parity constraint (in general) uses the sensitive variables $a$ both in estimating the conditional probabilities and for determining their thresholds \cite{Corbett2017, Lipton2018}. Therefore, solutions such as {\it disparate learning processes (DLPs)}, which allow the use of sensitive variables during model building but not prediction, are either sub- or equi-optimal \cite{Pedreshi2008, KamiranCalders2009, Kamiran2010, Kamishima2011, Zafar2017a}.

\section{Flavors of fairness definitions incorporating additional context} \label{causal_fairness} 
So far, we have discussed ``oblivious'' fairness based on summaries of the joint and marginal distributions of $Y$, $V$, $S$, and $D$. 
In this section, we consider fairness definitions that incorporate additional context, in the form of metrics and causal models, to inform fairness considerations.
This external context provides additional degrees of freedom for mapping social goals onto mathematical formalism.

\subsection{Metric fairness}\label{metric_fairness}

Assume there is a metric that defines similarity based on all variables, $m: \mathcal{V} \times \mathcal{V} \rightarrow \mathbb{R}$. Then, {\bf Metric Fairness} is defined such that  for every $v,v' \in \mathcal{V}$, their closeness implies closeness in decisions  $|\delta(v) - \delta(v')| \leq m(v,v')$. This is also known as {\it individual fairness}, the $m$-{\it Lipschitz property} \cite{Dwork2012}, and {\it perfect metric fairness} \cite{RothblumYona2018}. In cases where the metric only considers insensitive variables, $m(x,x')$, metric fairness implies unawareness.

Definitions of the metric differ. In the original work, Dwork et al. consider a similarity metric over {\it individuals} \cite{Dwork2012}. But in subsequent research, the metric is often defined over the {\it variables} input to the classifier \cite{RothblumYona2018,Kim2018b}. 

Either way, the metric is meant to ``capture ground truth" \cite{Dwork2012}. This inspired Friedler et al. to define the {\it construct space}, the variables on which we want to base decisions \cite{Friedler2016}. For example, suppose we want to base decisions on (the probability of) the outcome for an individual $i$. Let $I$ be a random individual drawn from the population. We can express the construct space as $\mathcal{CS} = \{ t_i \}$ where $t_i = P[Y = 1| I = i]$. But we cannot estimate $P[Y = 1| I = i]$ because we only have one individual $i$ and we do not observe their outcome in time to make the decision. Instead, we calculate scores $s_i = \psi(v_i)$ intended to estimate $P[Y=1|V=v_i]$. As noted above, the conditional probabilities $P[Y=1|V=v_i]$ are sometimes misleadingly called an individual's ``true risk" \cite{Corbett2017, CorbettGoel2018}. But the probabilities do not condition on the individual, only some measured variables. (The computer science literature sometimes conflates an individual with their variables, see e.g. \cite{Kearns2018}.)

Friedler et al. introduce an assumption they call WYSIWYG (what you see is what you get), i.e. that we can define a metric in the observed space that approximates a metric in the construct space: $m(v_i,v_{i'}) \approx m_{\mathcal{CS}}(t_i,t_{i'})$. To satisfy WYSIWYG, $m$ may need to be {\it aware} of the sensitive variables \cite{Dwork2012}. One reason is that the insensitive variables $X$ may predict $Y$ differently for different groups. For example, suppose we want to predict who likes math so we can recruit them to the school's math team. Let $Y=1$ be liking math and $X$ be choice of major. Suppose in one group, students who like math are steered towards economics, and in the other group towards engineering. To predict liking math, we should use group membership in addition to $X$.  Using this terminology, Dwork et al.'s metric can be defined as aligning differences in the construct space  with differences in the observed space \cite{Dwork2012}.

Friedler et al. also introduce an alternate assumption called WAE (we're all equal), i.e. that the groups on average have small distance in the construct space \cite{Friedler2016}. On this basis, we could adjust a metric in the observed space so that the groups have small distance \cite{Dwork2012}. Several papers describe methods adjusting the insensitive variables $X$ so that they are independent of group, which is consistent with adjusting the observed space so that it is consistent with a WAE understanding of the construct space \cite{JohndrowLum2017}. 

Though conceptually appealing, one major difficulty of {\it implementing} metric fairness is that it can be difficult to define the metric itself, especially in high dimensions. Recent work bypasses explicit elicitation of $m$ and instead queries individuals only on whether $|\delta(v) - \delta(v')|$ is small for many pairs of individuals $i$ and $i'$ \cite{jung2019eliciting}. For example, it requires data on whether individuals $i$ and $i'$ ought to be treated similarly without explicitly requiring the decision-maker to give an analytical expression for $m$. The objective function for model fitting then incorporates this notion of fairness by enforcing the elicited pairwise constraints.

\subsection{Causal definitions}
In this section, we discuss an alternative framework for conceiving of model fairness: causality. Causal definitions are a flavor of fairness, but we set it apart from Section \ref{flavors} given its different orientation.

Importantly, framing fairness issues with causal language can make value judgments more explicit. In particular, this framing allows practitioners to designate which causal pathways from sensitive attributes to decisions constitute ``acceptable'' or ``unacceptable'' sources of dependence between sensitive attributes and decisions. A number of the key questions involved in mapping social goals to mathematical formalism can thus be addressed by examining a causal graph, and discussing these value judgments.

 We have already touched on causal notions, considering the {\it potential} or {\it counterfactual} 
values under different decisions  in section \ref{goal} \cite{Rubin2005,ImbensRubin2015, HernanRobins2018}. Causal fairness definitions consider instead counterfactuals under different settings of a sensitive variable. Let $v_i(a)=(a,x_i(a))$ be the covariates if the individual had their sensitive variable set to $a$. We write $d_i(a) = \delta(v_i(a)) = \delta(a,x_i(a))$ for the corresponding decision, e.g. what would the hiring decision be if an black job candidate had been white? We define random variables $V(a), D(a)$ as values randomly drawn from the population. \cite{kohler2018eddie}

There is debate over whether these counterfactuals are well-defined. Pearl allows counterfactuals under conditions without specifying how those conditions are established, e.g. ``if they had been white". In contrast, Hern{\'a}n and Robins introduce counterfactuals only under well-defined interventions, e.g. the intervention studied by Greiner and Rubin: ``if the name on their resume were set to be atypical among black people" \cite{Greiner2011,HernanRobins2018}.

\begin{figure}[h!]
\center
\includegraphics[scale=.25]{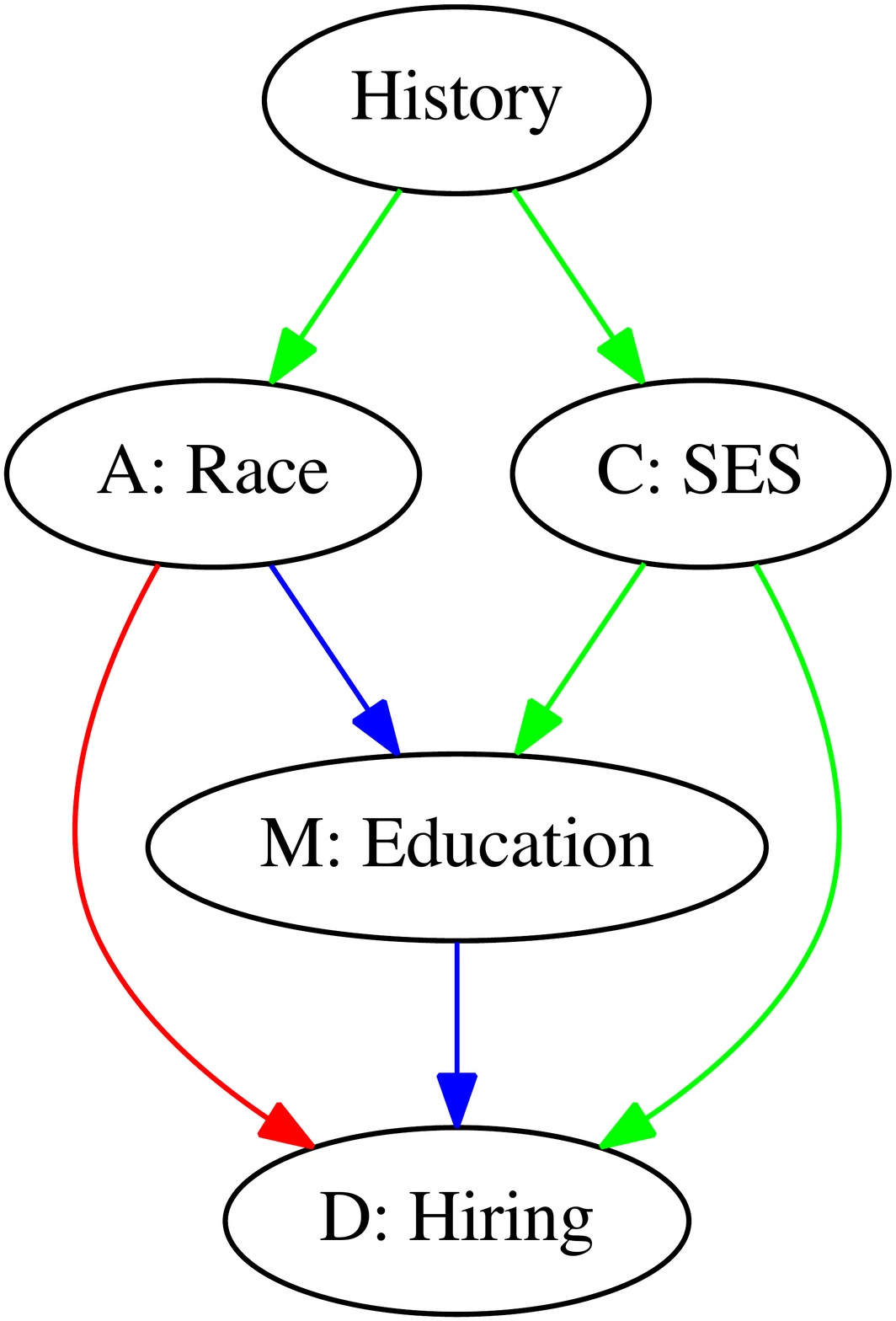}
\caption{A causal graph showing direct (red), indirect (blue), and back-door (green) paths from race to hiring.}\label{DAG}
\end{figure}

Putting these issues to the side, we can proceed to define fairness in terms of counterfactual decisions under different settings of a sensitive variable. 
Ordered from strongest to weakest, we have {\bf Individual Counterfactual Fairness}  ($d_i(a) = d_i(a')$ for all $i$), {\bf Conditional Counterfactual Fairness}\cite{Kusner2017} ($E[D(a)\ |\ \textrm{Data}] = E[D(a')\ |\ \textrm{Data}]$), and 
{\bf Counterfactual Parity}  ($E[D(a)] = E[D(a')]$). These first three causal definitions consider the {\it total} effect of $A$ (e.g. race) on $D$ (e.g. hiring). However, it is possible to consider some causal pathways from the sensitive variable to be fair. For example, suppose race affects education obtained. If hiring decisions are based on the applicant's education, then race affects hiring decisions. Perhaps one considers this path from race to hiring {\it through education} to be fair. It often helps to visualize causal relationships graphically, see Figure \ref{DAG}.\footnote{See Pearl \cite{Pearl2009} section 1.3.1 for a definition of causal graphs, which encode conditional independence statements for counterfactuals.}

In this cartoon version of the world, a complex historical process creates an individual's race and socioeconomic status at birth \cite{VanderWeele2014, JacksonVanderWeele2018}. These both affect the hiring decision, including through education. Let $x_i = (c_i,m_i)$ where $m_i$ are variables possibly affected by race, so $x_i(a) = (c_i,m_i(a))$. We can define effects along paths by defining fancier counterfactuals. Let $d_i(a',m_i(a))$ be the decision if applicant $i$ had their race set to $a'$, while their education were set to whatever value it would have attained if they had their race set to $a$ \cite{Nabi2018}. To disallow the red path in Figure \ref{DAG}, we can define {\bf No Direct Effect Fairness}: $d_i(a) = d_i(a', m_i(a))$ for all $i$.

However, $m_i(a)$ is only observed when $a_i = a$, so it is not possible to confirm no direct effect fairness without direct access to the model's inner workings. 

If we are willing to make stronger assumptions and assume ignorability, $M(a) \perp A\ |\ C$, we can, however,  check {\bf No Average Direct Effect Fairness}: $E[D(a)] = E[D(a', M(a))]$.

Beyond direct effects, one could consider other directed paths from race to be fair or unfair. For example, {\bf No Unfair Path-Specific Effects} specifies no average effects along unfair (user-specified) directed paths from $A$ to $D$ \cite{Nabi2018}. Relatedly, {\bf No Unresolved Discrimination} states that there should exist no directed path from $A$ to $D$ unless through a {\it resolving variable} (a variable that is influenced by $A$ in a manner that we accept as nondiscriminatory)  \cite{Kilbertus2017}. 

All of the above causal definitions consider only directed paths {\it from} race. In Figure \ref{DAG}, these include the red and blue paths. But what about the green paths? Known as {\it back-door paths} \cite{Pearl2009}, these do not represent causal effects of race, and therefore are permitted under causal definitions of fairness. However, back-door paths can contribute to the association between race and hiring. Indeed, they are why we say ``correlation is not causation."\footnote{If $C$ satisfies the {\it back-door criterion} ($C$ includes no descendants of $A$ and blocks all back-door paths between $A$ and $D$) in a causal graph, then unconfoundedness ($D(a) \perp A | C\ \forall a$) holds \cite{Pearl2009, perkovic2015complete}. The converse is not true in general.} Zhang and Bareinboim decompose the total disparity into disparities from each type of path (direct, indirect, back-door) \cite{ZhangBareinboim2018}. In contrast to the causal fairness definitions, {\it health disparities} are defined to include contribution from back-door paths (e.g. through socioeconomics at birth) \cite{IOM2003, JacksonVanderWeele2018}. 

Causal definitions of fairness focus our attention on how to compensate for causal influences at decision time. Causal reasoning can be used instead to design interventions (to reduce disparities and improve overall outcomes), rather than to define fairness. In particular, causal graphs can be used to develop interventions at earlier points, prior to decision-making \cite{JacksonVanderWeele2018,Barabas2018}.

\section{Ways Forward}\label{ways_forward}

In this paper, we have been careful to identify assumptions, choices, and considerations in prediction-based decision-making that are and are not challenged by various mathematically-defined notions of fairness. 
This paper does not, however, discuss what to {\it do} with a flavor of fairness.
The dominant focus on the Fairness in ML literature has been to constrain decision functions to satisfy particular fairness flavors, and to treat fair decision-making as a constrained optimization problem (see, e.g., \cite{Hardt2016}).

Another way forward is to address the choices and assumptions outlined in Section \ref{preliminaries} directly. Here we sketch that approach, including pointing to some of the relevant statistics literature.

Starting with clearly articulated goals can improve both fairness and accountability. To best serve those goals, consider whether interventions should be made at the individual or aggregate level. Carefully describe the eligible population to clarify who is impacted by the possible interventions. Expanding the decision space to include less harmful and more supportive interventions can benefit all groups and mitigate fairness concerns.

To build a decision system aligned with the stated goals, choose outcomes carefully, considering data limitations. Using {\it prior information} \cite{Gelman2013} can help specify a realistic utility function. For example, instead of assuming benefits and harms are constant across decisions (the ``symmetric'' assumption), prior data can inform a more realistic distribution. 

Instead of assuming one potential outcome is known, causal methods can be used to estimate effects of decisions. Furthermore, these effects may not be separate and constant across the population. As such, causal methods can be used to study interference \cite{ogburn2014causal,miles2017} and heterogeneous effects \cite{gelman2015connection}. 

Documenting data collection (e.g. sampling and measurement) \cite{gebru2018datasheets, holland2018dataset} enables modeling that appropriately accounts for the specific conditions under which data has been collected (Chapter 8 of \cite{Gelman2013}, \cite{rubin1976inference,little2004model}). Combining all choices in one expanded model \cite{yao2018using} can mitigate sensitivity of decisions to model selection. Documenting model performance \cite{ Stoyanovich2018, yang2018nutritional, mitchell2018model}, both overall and within subgroups, is crucial to effective evaluation and can also help check decision systems against some of the ``fairness" definitions from Section \ref{flavors}.

\section{Conclusion}\label{conclusion}

The fact that much of the work on fairness in prediction-based decision making has emerged from computer science and statistics has sometimes led to the mistaken impression that these concerns only arise in cases that involve predictive models and automated decision making. Yet many of the issues identified in the scholarship apply to human decision making as well, to the extent that human decision making also rests on predictions. Notably, scholars have emphasized that the tradeoff between different notions of fairness would apply even if a human were the ones making the prediction \cite{Kleinberg2016}. Rejecting model-driven or automated decision making is not a way to avoid these problems. 

At best, formal fairness metrics can instead illustrate when changes to prediction-based decision making are insufficient to achieve different outcomes---and when interventions are necessary to bring about a different world. Recognizing the trade-offs involved in prediction-based decision making does not mean that we have to accept them as a given; doing so can also spur us to think more creatively about the range of options we have beyond making predictions to realize our policy or normative goals.

Recent criticisms of this line of work have rightly pointed out that quantitative notions of fairness can funnel our thinking into narrow silos where we aim to make adjustments to a decision-making process, rather than to address the structural conditions that sustain inequality in society \cite{GreenHu2018, Ochigame2018}. While algorithmic thinking runs such risks, quantitative modeling and quantitative measures can also force us to make our assumptions more explicit and clarify what we're treating as background conditions (and thus \textit{not} the target of intervention). In doing so, we have the opportunity to foster more meaningful deliberation and debate about the difficult policy issues that we might otherwise hand wave away: what is our objective and how do we want to go about achieving it?

Used with care and humility, the recent work on fairness can play a helpful part in revealing problems with prediction-based decision making and working through addressing them.  While mathematical formalism cannot solve these problems on its own, it should not be dismissed as necessarily preserving the status quo. The opportunity exists to employ quantitative methods to make progress on policy goals \cite{MODA,potash2015predictive}.  The fairness literature that we have reviewed facilitates a critical reflection on the way we go about choosing those goals as well as procedures for realizing them.

\section*{Acknowledgements}
Jackie Shadlen provided substantial insights that contributed greatly to this paper. We are grateful to all cited authors for their work and for answering our many questions. Funding for this research was provided by the Harvard-MIT Ethics and Governance of Artificial Intelligence Initiative. 

\bibliographystyle{plain}
\bibliography{fairness_catalogue}

\end{document}